# Morphology and Mechanical Properties of Nanocrystalline Cu/Ag Alloy

Ao Li[1], Izabela Szlufarska[1,2]
[1]Materials Science Program, University of Wisconsin, Madison, WI
[2]Department of Materials Science and Engineering, University of Wisconsin, Madison, WI

**Abstract:**
Hybrid Monte Carlo (MC)/molecular dynamics (MD) simulations are conducted to study the microstructures of nanocrystalline (nc) Cu/Ag alloys with various Ag concentrations. When the Ag concentration is below 50 Ag atoms/nm², an increase in Ag concentration leads to a gradual growth of monolayer grain boundary (GB) complexions into nanolayer complexions. Above the concentration of 50 Ag atoms/nm², wetting layers with a bulk crystalline phase are observed. The effects of Ag on mechanical properties and deformation mechanisms of nc Cu/Ag alloys are investigated in MD simulations of uniaxial tension. GB sliding resistance is found to first increase and then decrease with an increase in Ag concentration. Surprisingly, we also find that the dislocation density decreases monotonically with an increase in Ag concentration, which suggests that the grain interiors are softened by the introduction of Ag dopants at GBs. In addition, there is a critical Ag concentration that maximizes flow stress of nc Cu/Ag alloys. The flow stress, GB sliding resistance, and the intragranular dislocation densities become less sensitive to Ag dopants when the grain diameter increases from 5nm to 40nm.

## 1. Introduction

Because of their high electrical conductivity, Cu alloys are of technological interest in a number of applications, such as high-field pulsed magnets.[1] These applications often require a relatively high mechanical strength – higher than the strength of pure polycrystalline Cu. One way to enhance mechanical properties of Cu is by grain refinement to the nanometer regime, which has been reported to lead to significant increases in strength[2], hardness[3], and ductility[2,4]. Both experimental[2,3,4] and simulation[5,6,7] studies have shown that there is an optimum grain size that maximizes the strength of nanocrystalline (nc) Cu. It was proposed that this maximum strength is due to a crossover of the dominant deformation mechanisms, which are dislocation-mediated plasticity in the large grain size regime and GB sliding in the small grain size regime.[5,6,7] Another promising way to strengthen Cu is through alloying. In particular Cu/Ag is a widely studied material[8,9,10] not only because it is a simple model alloy system, but also because it has both high strength and high electrical conductivity. Earlier studies showed that the ultimate tensile strength of polycrystalline Cu/Ag alloys with micrometer-sized grains increases when Ag concentration increases from 0 to 6%.[11] In another study of polycrystalline Cu/Ag alloys, Sakai *et al* explored a large Ag concentration regime and found that increasing Ag concentration from 2% to 60% leads to initial strengthening of the alloy with Ag concentration followed by saturation when Ag concentration reaches 10-20%.[1] In this study, Cu/Ag alloy samples were cold drawn and heat treated, and the strength was improved by formation of precipitates of Ag and Cu in the solid solution matrix.

Several mechanisms have been proposed to explain how Cu is strengthened by alloying with Ag. In one of these mechanisms Ag dopants reduce the GB energy and make GB sliding more difficult, which in turn increases the yield strength of the nc alloy.[12] Ag dopants can also stabilize the structure of nc Cu (i.e., suppress the grain growth) and thereby preserve the advantageous mechanical properties of nc Cu.[13] Finally, when the concentration of Ag is supersaturated in the Cu matrix, Ag precipitates can nucleate and grow in the grain interiors, which may strengthen Cu by impeding dislocation motion.

The activation of the aforementioned strengthening mechanisms is highly dependent on the morphology of Cu/Ag alloys and the distribution of Ag in the alloys, i.e., whether Ag dopants are located at GBs or precipitated in grain interiors. The desirable microstructure can be achieved by controlling synthesis conditions. For instance, Cu/Ag alloys prepared by casting and by physical vapor deposition have entirely different morphologies. Cu/Ag alloys casted from pure Cu and pure Ag are composed of Cu-rich grain interior regions and Ag-rich GB regions.[14] When the concentration of Ag is low, Ag phase is dispersed as isolated islands at GBs. In the large Ag concentration regime, Ag phase shows a continuous network of layered regions at GBs. On the other hand, Cu/Ag alloys prepared by physical vapor deposition and subsequent annealing have Ag grains comparable in size to Cu grains, although formation of GB wetting layers of Ag can also be achieved in this synthesis method by adding tungsten.[15]

Not only the type of the synthesis technique but also the details of the synthesis process can have a strong influence on the microstructure. For example, the dominant types of Cu/Ag interfaces formed during casting can be controlled by setting the furnace removal rate. When the furnace removal rate is as low as 0.46 mm/h, only Cu/Ag interfaces with cube-on-cube orientations are observed. However, when the furnace removal rate is as high as 76 mm/h, incoherent-twin interfaces with the $\{\bar{3}13\}_{Ag}||\{\bar{1}12\}_{Cu}$ habit plane comprise over 66% of all interfaces in the sample.[16] Altering the fractions of cube-on-cube and incoherent twin interfaces in the Cu/Ag alloys by changing processing parameters can significantly influence mechanical properties and deformation mechanisms of the alloy.[16,17,18] The cube-on-cube interfaces are known to be weak barriers to transition of twinning partials, whereas the incoherent twin interfaces are strong barriers to twinning partials. The difference in the ability of different Cu/Ag interfaces to transfer twinning partials explains some of the details of the microstructures of the Cu/Ag alloys. For instance, for both cube-on-cube and incoherent twin interfaces, twinning partials are first nucleated in the Ag phase. Since the cube-on-cube interfaces are able to transfer the twinning partials, twinning partials glide through the interfaces and lead to twinning in the Cu phase. However, the incoherent twin interfaces are not able to transfer twinning partials and twins are not observed in the Cu phase.[17] Moreover, plastic strain recovery has been observed in the case where the majority of interfaces are the incoherent twin interfaces, whereas no plastic strain recovery has been observed when only cube-on-cube interfaces are present.[18]

In summary, mechanical properties of Cu can be improved by grain refinement to the nanometer length scale and/or by alloying. Studies have been dedicated to understanding how processing conditions control the microstructure and the resulting mechanical properties of Cu/Ag alloys with critical dimensions in the range of hundreds of nanometers to tens of microns. However, limited studies have been conducted in achieving enhanced mechanical properties through a combination of alloying and of grain refinement to the regime of tens of nanometers and smaller. In one recent study, the yield strength of nc Cu/Ag alloy has been investigated and shown to increase due to Ag addition, but only with up to 1% Ag concentration.[12] It is unknown whether this trend will persist for larger Ag concentrations. In the same study the authors found that Ag dopants at GBs influence GB sliding by changing GB energy,[12] however the possible influence of GB dopants on dislocation-mediated plasticity in grain interiors has not been explored even in the low Ag concentration regime.

## 2. Methods

## 2.1 Preparation of nc Cu samples

We conduct MD simulations of nc Cu/Ag alloys in the range of overall concentration between 0 and 24.4%. MD simulations are performed with the LAMMPS software package[19] using the embedded atoms method (EAM) force field[20,21]. The Cu/Ag EAM potential was shown to correctly reproduce such properties as free energy of solvation, stacking fault energy, and twinning energy.[20,21] This potential has also been used in MD simulations of impurity segregation[12] and interface-mediated mechanical properties[18], and both of these studies showed results consistent with theoretical predictions and experimental observations. We first generate nc structures of pure Cu using the Voronoi algorithm. The average grain diameters in different samples are 5, 10, 15, 20.8, 30, 40, and 50nm and they correspond to 24.4, 25.4, 25.5, 25.6, 45.6, and 74.4 million atoms, respectively. Grain diameter distribution follows the Gaussian distribution, which is consistent with experimental distributions reported for nc Cu.[22] Periodic boundary conditions are applied to all three dimensions. Samples prepared in this way are then relaxed for 1ns at 300K and zero pressure using Berendsen's thermostat and barostat.

## 2.2 Alloying

After relaxation, nc Cu samples with 5, 15, and 40nm grain diameters are selected for alloying with various concentrations of Ag. A hybrid MC/MD method[23] is utilized to introduce Ag dopants to the nc Cu samples, and to relax the structures energetically. The MC/MD simulations are performed at 300K and zero pressure for the total of time of 1ns. The key idea of this method is to perform iteratively transmutational MC steps, which sample the semi-grand-canonical ensemble, and MD simulations for 50fs. In the MC run, trial swaps are carried out by randomly selecting an atom (which can be either Cu or Ag) and by replacing it with the other species. The trial moves are accepted or rejected according to the Metropolis algorithm, which calculates acceptance probability $A_s$ based on the energy change $\Delta U$ during atom swaps according to the following formula $A_s = \min\{1, \exp[-\frac{(\Delta U + \Delta \mu N \Delta c)}{k_B T}]\}$. In this expression, $k_B$ is the Boltzmann constant, $T$ is the temperature, $\Delta \mu$ is the chemical potential difference between the two elements, and $N$ is the total number of atoms in the system. Here, $\Delta \mu$ is only an arbitrary parameter, which does not correspond to any physical reference state. This parameter controls the acceptance probability $A_s$ and it is changed to achieve the desired final concentration after a fixed number of hybrid MC/MD steps. This value of $\Delta \mu$ is used only to prepare Cu/Ag alloy samples in a computationally efficient manner (i.e., without having to model the synthesis itself) and this $\Delta \mu$ is not used anywhere else in this paper. During the 1ns alloying and relaxation, there are 20,000 trial moves on the average performed on each atom in the system. One example of the doped nc Cu/Ag alloy is shown in Fig.1a. As expected, Ag dopants are predominately segregated to GB regions because Cu/Ag is an immiscible alloy system. Since Ag dopants are segregated to the GBs, it is convenient to calculate Ag concentrations at GBs as $C_{Ag} = \frac{N_{Ag}}{A}$, where $N_{Ag}$ is the total number of Ag atoms in the system, and $A$ is the total GB area in the sample. The GB area is calculated using the Voronoi tessellation method.

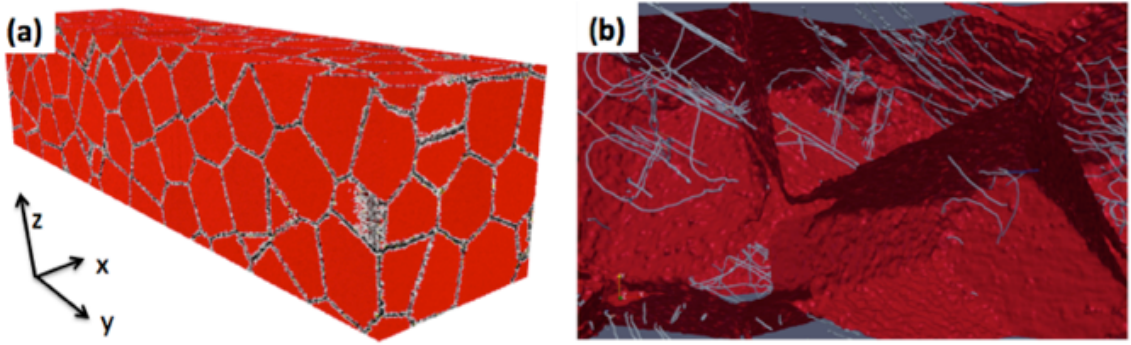

*Figure 1.* (a) Ag-doped nanocrystalline Cu with 15 nm average grain diameter and 24.9 Ag atoms per nm² concentration. Red atoms are FCC Cu atoms. White atoms are defect Cu atoms, which means that they do not have FCC, HCP, or BCC structures. Ag atoms are colored black. Defect Cu atoms and Ag atoms mark the GB regions, and FCC Cu atoms are primarily found in grain interior. (b) Intragranular dislocations (grey). Red planes represent GBs.

**2.3 Loading**

Uniaxial tension simulations are performed at 300 K with engineering strains up to 15% and with a strain rate of $5 \times 10^8$ s$^{-1}$. The deformation strain is applied along the z-direction through alternating steps of straining the system by 0.05% and 1ps MD equilibration. During the entire processes, the Nose-Hoover thermostat is used in x-, y-, and z-directions to keep the temperature at 300K. The Nose-Hoover barostat is used to maintain zero pressure in x- and y-directions, and the x- and y- dimensions of the system are free to change to account for the Poisson effect. Flow stress is defined as the average value of stress measured between the engineering strain of 12% and 15%. In this regime the stress fluctuates around an approximately constant value.

**2.4 Analysis Methods**
**2.4.1 Intragranular Dislocation Density**

Dislocation Extraction Algorithm (DXA)[24] is used to identify atoms that belong to dislocations and to calculate the total dislocation length in deformed samples. An example of a dislocation network identified with this method is shown in Fig.1b. Intragranular dislocation density is calculated in four steps. In the first step, GB atoms are identified using the common neighbor analysis (CNA).[25] Specifically, the CNA algorithm is used to identify the local crystallographic order of each copper atom, i.e., atoms are identified as having FCC, hexagonal closed packed (HCP), or body centered cubic (BCC) structure. Atoms that do not have any of these local structures are labeled as *defect atoms*. As GB regions are comprised of Cu defect atoms and Ag atoms (since Ag atoms segregate to GBs), we use these two criteria to identify GB regions. In the second step, atoms identified as belonging to GBs are deleted. In the third step, the structures without GB atoms are analyzed by DXA, which calculates the total length of intragranular dislocations $l_{interior}$. In the last step, intragranular dislocation density $\rho$ is calculated as $\rho = \frac{l_{interior}}{V_{interior}}$, where $V_{interior}$ is the total volume of the grain interiors. $V_{interior}$ is calculated as $(N - N_{GB}) \times v_{Cu}$, where $N$ is the total number of atoms in the system, $N_{GB}$ is the number of deleted GB atoms, and $v_{Cu}$ is the volume of a copper atom in pure FCC lattice. $v_{Cu}$ is calculated by dividing the volume of a unit cell in a Cu FCC lattice by the number of atoms in the unit cell.

### 2.4.2 Atomic local shear strain

To investigate deformation mechanisms and quantify plastic deformation, we calculate the atomic local shear strain.[26] Two atomic configurations are required to calculate the atomic local shear strain, a current configuration and a reference configuration. Here, the configuration of a deformed system is set as the current one and the initial configuration before deformation is set as the reference one. For each atom $i$ in the system, its position relative to its neighbors changes after deformation. This position change can be represented as $\{d_{ji}^0\} \to \{d_{ji}\}, \forall j \in N_i^0$, where $d_{ji}^0$ and $d_{ji}$ are three-dimensional vectors representing the distance between atoms $j$ and $i$ in the reference configuration and the current configuration, respectively. $j$ atom is one of the nearest neighbors of atom $i$, and $N_i^0$ is the total number of the nearest neighbors of atom $i$. A local transformation matrix $J_i = (\sum_{\forall j \in N_i^0} d_{ji}^{0T} d_{ji}^0)^{-1} (\sum_{\forall j \in N_i^0} d_{ji}^{0T} d_{ji})$ can be calculated based on $d_{ji}^0$ and $d_{ji}$, so that $J_i$ minimizes the expression $\sum_{\forall j \in N_i^0} |d_{ji}^0 J_i - d_{ji}|^2$. A local strain tensor $\eta_i$ is then calculated from $J_i$ using the formula $\eta_i = \frac{1}{2}(J_i J_i^T - I)$. $\eta_i$ quantifies the atomic-scale local deformation around atom $i$. Von Mises shear strain invariant is subsequently calculated from the local strain tensor as $\eta_i^{Mises} = \sqrt{\eta_{yz}^2 + \eta_{xz}^2 + \eta_{xy}^2 + \frac{(\eta_{yy}-\eta_{zz})^2+(\eta_{xx}-\eta_{zz})^2+(\eta_{xx}-\eta_{yy})^2}{6}}$. $\eta_i^{Mises}$ is assigned to each atom and it is referred to as the atomic local shear strain. Analysis of the atomic local shear strain is used in this paper to illustrate regions where plastic deformation is concentrated.

### 3. Results and Discussion
#### 3.1 Microstructure

Figure 1a shows that almost all Ag dopants are segregated to the GB regions of the Cu/Ag alloy after the hybrid MC/MD alloying and equilibration, which is reasonable given that it is an immiscible binary alloy. In order to identify the GB complexions for different concentrations, cross-sectional views of Cu/Ag alloy with different concentrations are shown in Fig.2. Specifically, Fig.2a shows the cross-sectional view of Cu/Ag alloy with 15nm grain diameter and Ag concentration of 1.88 Ag atoms/nm$^2$ (0.55% overall concentration). The GB between grains marked as G1 and G2 is a high-angle tilt GB. In this type of GB, Ag atoms are isolated from each other and form a monolayer complexion of dopants. Similar monolayer complexion of GB dopants has also been observed experimentally in other eutectic binary alloys with low dopant concentrations, such as Bi-doped bicrystal Cu.[27]

When Ag concentration increases, Ag atoms are no longer isolated from each other (Fig.2a) and small Ag precipitates begin to form at GBs (Fig.2b). With a further increase of Ag concentration, these precipitates have a tendency to grow along GBs, rather than into grain interiors. The isolated Ag precipitates gradually become connected to each other and form nanolayer complexions, which are several atomic layers thick (i.e., 1-3nm). So far, the isolated Ag precipitates and nanolayer complexions are still regarded as part of the nc Cu GBs. They lack long-range crystalline order and are not considered as a separate phase. It is also worth mentioning that in the case of general GBs studied here we did not observe a discontinuous jump in GB thickness with increasing dopant concentration, whereas such discontinuity has been previously reported for high-symmetry GBs with ordered structures.[27] This is because for high-symmetry GBs with ordered structures, the energies of GB sites are distributed discretely. Dopants will occupy the GB sites layer by layer, which leads to the discrete thickening at low dopants concentrations. In contrast, for a general GB, the thickness of GB complexions increases gradually. The reason is that GB structures are not ordered and the energies of sites have a more random (continuous) distribution. As a result, when Ag dopants fill the continuously distributed energy states, the thickness of the GB complexions increases gradually.

While Ag precipitates and nanolayer complexions are still constituents of GBs in nc Cu, wetting layers, which have bulk Ag phase with FCC crystalline structures, begin to develop when Ag concentration is above a certain concentration. This threshold concentration is around 50 Ag atoms/nm$^2$, which for nc Cu/Ag alloy with 40nm average grain diameter corresponds to 5.5% overall Ag concentration. As shown in Fig.2c, when $C_{Ag}$ is as high as 223.7 Ag atoms/nm$^2$ (corresponding to 24.4% overall concentration in an alloy with 40 nm grain diameter), Ag atoms form wetting layers along some of the GBs. First, it is important to point out that not all types of GB complexions transform into wetting layers. To understand which GB types are more prone to this transformation, we investigate structures of the newly formed Cu/Ag interfaces. The atomic details of two major types of Cu/Ag interfaces are shown in Fig.2d and Fig.2e, which are cube-on-cube interfaces with $\{111\}_{Ag}||\{111\}_{Cu}$ habit plane and hetero-twin interfaces with $\{111\}_{Ag}||\{111\}_{Cu}$ habit plane, respectively. The cube-on-cube and hetero-twin interfaces have been reported to be the predominant interfaces in cast two-phase Cu/Ag alloys.[18,28,29] It can therefore be reasonably expected that their energy is lower than general types of Cu/Ag interfaces. As a result, when a newly formed Ag phase in the Cu GB region is able to form the aforementioned low-energy interfaces, this GB region becomes preferable for formation of Ag wetting layers. We found many defects, such as dislocations, stacking faults, and twins, to be present inside the newly formed Ag phases. Dislocations are nucleated during the MD equilibration and they relieve the interfacial stress due to lattice mismatch between Ag and Cu phases. The dislocation activity inside Ag layers changes orientation relations between Ag and Cu lattices[17,18] enabling formation of cube-on-cube or twin interfaces. Similar interfacial dislocations have been reported for the cube-on-cube and hetero-twin interfaces observed in the as-cast Cu/Ag system in experiments.[18,29]

It is interesting to ask why the newly formed Ag phase has a shape of a wetting layer, instead of a spherical one since the spherical shape minimizes the Cu/Ag interfacial area. This observation, too, can be explained by the low energies of cube-on-cube and hetero-twin interfaces. Even though a Ag phase with isotropic shape can reduce the interfacial area, the total energy is high because the interface is comprised of segments of general Cu/Ag interfaces, which typically have a higher energy. Our simulations show that the Ag phase has the tendency to maximize the area of low-energy cube-on-cube and hetero-twin interfaces, which leads to the observed layered morphology. The layered structures with predominately cube-on-cube and hetero-twin interfaces have been previously reported for cast Cu/Ag alloys.[28,29]

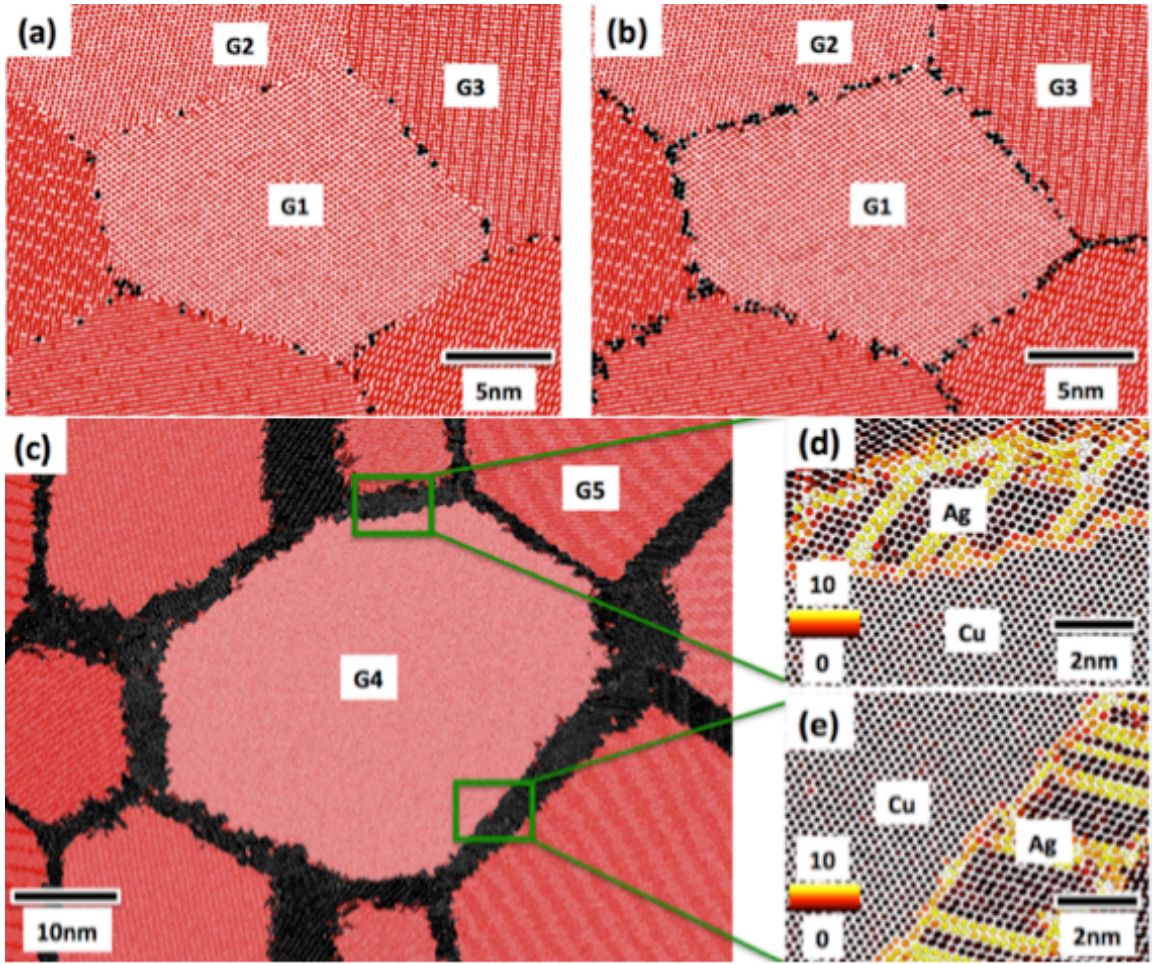

*Figure 2. Cross-sectional view of Cu/Ag microstructures with different Ag concentrations. (a) and (b) Samples with 15nm grain diameter and Ag concentration of 1.88 Ag atoms/$nm^2$, 8.35 Ag atoms/$nm^2$, respectively. (c) Grain diameter and Ag concentration are 40nm and 223.7 Ag atoms/$nm^2$, respectively. (d) and (e) are two enlarged local structures of the regions marked by rectangles in (c). These structures have cube-on-cube and hetero-twin Cu/Ag interfaces in (d) and (e), respectively. In (a)-(c), red color represents Cu and black represents Ag. In (d) and (e), atoms are colored by the centrosymmetry parameter. When the centrosymmetry parameter of an atom is close to 0, the atom has FCC local structure, while a larger centrosymmetry parameter indicates defects of dislocation cores, stacking faults, twins and interfaces. We used gradient coloring, which means that the local structure deviates from the perfect FCC structure more as the color changes gradually from black to white. Ag atoms are also visualized to be slightly larger than Cu atoms.*

GB energy is calculated using the formula $\frac{E_{tot} - e_{Ag} \times N_{Ag} - e_{Cu} \times N_{Cu}}{A}$, where $E_{tot}$ is the total energy of the system, $e_{Ag}$ is the cohesive energy of FCC Ag crystal, $N_{Ag}$ is the number of Ag atoms in the Cu/Ag alloy, $e_{Cu}$ is the cohesive energy of FCC Cu crystal, $N_{Cu}$ is the number of Cu atoms in the alloy, and $A$ is the total area of GBs. In Fig.3a we plot GB energies in the Cu/Ag alloy with 15nm grain diameter for a range of Ag concentration below the threshold values at which Ag begins to form a bulk phase (a wetting layer) along the GBs. We find that the GB energy has a minimum for a certain Ag concentration. One should keep in mind that the exact concentration corresponding to the minimum of energy depends on the reference states used in the GB energy calculations. Such reference state energies are highly dependent on the synthesis details and here we use the cohesive energies $e_{Cu}$ and $e_{Ag}$ of the single crystal Cu and Ag, respectively.

There are two competing effects of Ag dopants on the GB energy. The first one is that Ag dopants can reduce the GB energy by reducing the local stress at GBs. The initial decrease in GB energy with Ag concentration is due to the fact that there are sites at Cu GBs with sufficiently high local energies so that this energy will decrease if Ag dopants are segregated there. Such sites may include those with a relatively large GB free volume and because Ag atoms are larger than Cu, filling the sites with Ag dopants may reduce the tensile stress of these sites. On the other hand, Ag dopants can also increase the GB energy by making the GB thicker (increasing the total number of GB atoms). This trend is shown in Fig.3b. As a result, the volume fraction of the disordered GB regions increases and the total energy of the system increases as well. In the low Ag concentration regime, the dominant effect of Ag dopants is reduction of GB energy by releasing some of the local stress. Above a critical concentration of around 10-15 Ag atoms/nm², the GB energy begins to increase and the effect of Ag dopants on thickening the GBs dominates. The critical concentrations are comparable for different grain sizes if we use the units of the number of Ag atoms per GB area. However, the critical overall concentrations for different grain sizes are different since the volume fraction of GBs in samples with different grain sizes are not the same. The critical overall concentrations for 5nm, 15nm and 40nm samples are 14%, 3.5%, and 1.2%, respectively.

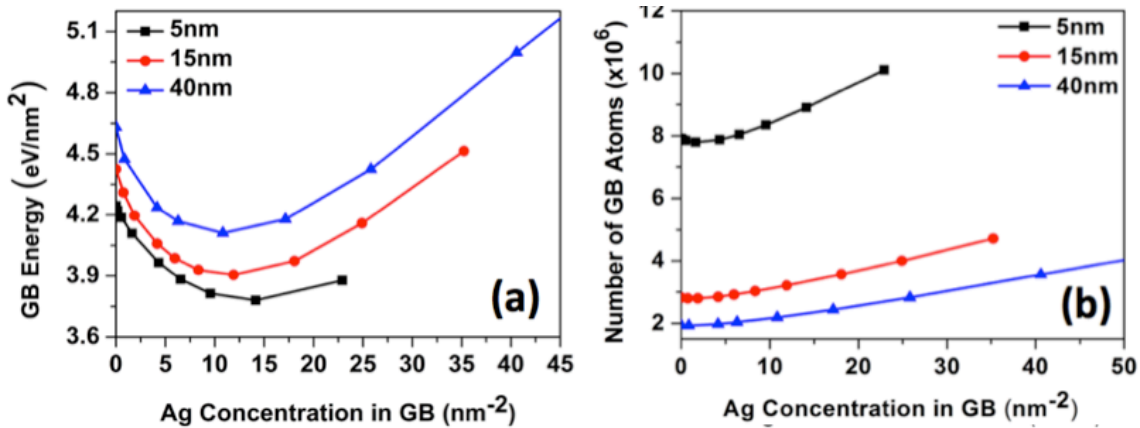

**Figure 3.** Dependence of (a) GB energy, and (b) the total number of GB atoms (representative of GB thickness) on Ag concentrations for samples with the average grain diameters of 5nm, 15nm, and 40nm.

### 3.2 Deformation Mechanisms
#### 3.2.1 Stress-Strain Curve of Pure nc Cu

We now turn to the questions of how Ag dopants influence the mechanical properties and deformation mechanisms of nc Cu/Ag alloys. Before analyzing the Cu/Ag alloy, it is instructive to briefly discuss mechanical response of pure nc Cu. Figure 4a shows a stress-strain curve of a pure nc Cu sample with a 15nm average grain diameter during uniaxial tension. There are three strain regimes corresponding to different deformation mechanisms. The elastic regime extends to the engineering strain of 1.8%. The Young's modulus is determined as the slope of a linear fit of the stress-strain function in the strain range between

0 and 0.3%. The second regime, extending from 1.8% to 3.7% strain, is the GB sliding regime. The 0.2% offset stress (used here as the definition of the yield strength) is 1.64GPa. It corresponds to the 1.8% total engineering strain. Beyond this strain the sample begins to deform plastically and the deformation is localized in GBs. This localization is evidenced by the analysis of the atomic local shear strain $\eta_i$ in Fig.4b. We find that GB sliding occurs by atomic shuffling due to stress-assisted free volume migration at GBs and that GB sliding is accommodated by free volume-mediated migration of GBs and triple junctions. These mechanisms have been previously reported to be dominant in nc metals at modest temperatures and they have been described in detail in Ref.[30] Because in the early stages of plasticity intragranular dislocation activity is negligible, the 0.2%-offset yield stress is a measure of the resistance of GBs to sliding. Earlier studies of uniaxial compression of ultrananocrystalline diamond (where GB sliding is the only active deformation mechanism for a large range of strains) have shown directly that the yield strength is linearly dependent on the GB shear strength.[31] In the case of uniaxial compression of the ultrananocrystalline diamond, the ratio between the yield stress and the GB shear strength was found to be 2.00±0.28GPa, but this value may be different for nc Cu.

At the strain of 3.7%, dislocations are nucleated and the stress reaches a maximum value. Beyond this strain threshold, plastic deformation is mediated by both, GB sliding and dislocation activity (Fig.4c). The stress decreases in this regime and almost reaches a plateau at the strain of 12%. Flow stress is calculated as the average value of stress in the strain range between 12% and 15% and it is equal to 1.99GPa. The fact that the stress shows a slight decrease in the strain range between 12% and 15% does not affect the results presented in this paper. Specifically, in the following discussion we will show how Ag dopants influence the yield stress and the flow stress of the nc Cu/Ag alloy. We found that the effect of Ag doping is one order of magnitude larger than the drop in the stress-strain curve shown in Fig.4a. Additionally, since all the samples show the same small decrease in stress in the large strain regime, this effect is cancelled out and therefore can be ignored when we compare flow stress of samples with different Ag concentrations.

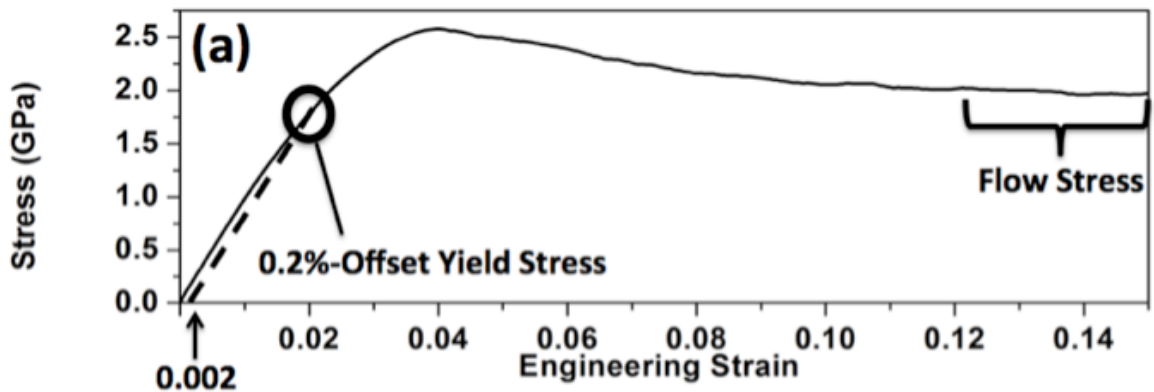

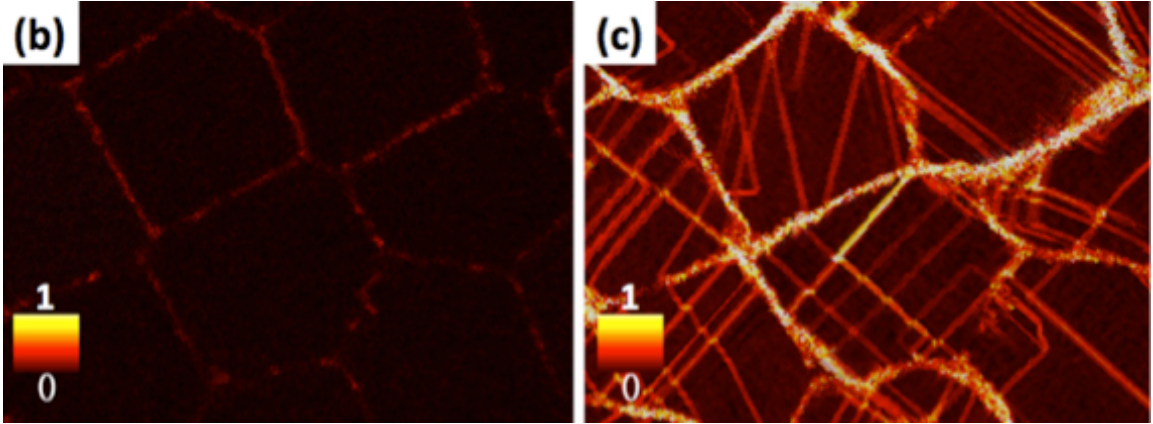

*Figure 4. The stress-strain curve of nc Cu with the average grain diameter of 15nm (a). In (a), the dashed line with the slope equal to the Young's modulus is added to determine the 0.2% off-set yield stress (marked with a circle). (b) and (c) Cross-sectional views of the sample at the engineering strain of 2% and 15%, respectively. Atoms in (b) and (c) are colored by the values of the calculated atomic local shear strain $\eta_i$.*

### 3.2.2 0.2%-Offset Yield Stress

The yield stress as a function of Ag concentration for nc Cu samples with the average grain sizes of 5, 15, and 40 nm is shown in Fig.5a. As discussed earlier, this yield stress is expected to scale linearly with the stress required for GB sliding (the GB sliding resistance). The yield strength has a maximum at a finite Ag concentration and this maximum is more pronounced for samples with smaller grain sizes. Here, we only consider Ag concentrations for which wetting layers of the Ag phase have not yet been formed. That means that all samples can be viewed as nc Cu with Ag dopants at GBs. In order to understand the trend in the yield strength (or equivalently in the GB resistance to sliding) shown in Fig.5a, we consider two possible strengthening mechanisms of Ag dopants. The first mechanism is the effect of Ag dopants on the strength of chemical bonds in the GB region. However, this effect can be excluded based on the following argument. The cohesive energy of FCC Cu is -3.5eV and the cohesive energy of FCC Ag is -2.8eV. Regarding the cohesive energy of an FCC metal as the energy required to break 12 bonds of an atom with its nearest neighbors, the bond energy of Cu is $e_{Cu}$ = 0.29eV and the bond energy of Ag is $e_{Ag}$ = 0.23eV. The energy of a Cu-Ag bond can be approximated as $e_{Cu/Ag} = \sqrt{e_{Cu} \times e_{Ag}}$=0.26eV. This analysis shows that introducing Ag dopants to Cu GBs will lower the bond energy and therefore the stress required to break chemical bonds. Consequently if this chemical bond effect on GB sliding was dominant, the GB sliding resistance would decrease due to Ag doping in the small Ag concentration regime, but the opposite trend is observed in our simulations. The second effect of Ag we consider here is the impact of such dopants on the GB free volume, because stress-driven migration of free volume is known to mediate GB sliding.[30] Since GBs are more disordered than grain interiors, there exist free volume in the vicinity of GB atoms. During deformation, local stress in GBs may lead to local shearing, which is mediated by atomic jumps or by free volume migration. These local events collectively mediate GB sliding. It can be reasonably assumed that at a given stress level, the larger the free volume, the easier the local shearing around this atom. In order to quantify how the free volume in GBs evolves with Ag concentration, we use $k$ to be the normalized excess free volume per atom, defined as the ratio between the excess free volume of GB atoms and the reference volume of the same number of atoms in a perfect crystal. Specifically, $k = \frac{V_{GB} - N_{Cu,GB} \times v_{Cu} - N_{Ag,GB} \times v_{Ag}}{N_{Cu,GB} \times v_{Cu} + N_{Ag,GB} \times v_{Ag}}$, where $V_{GB}$ is the total volume of GB atoms, $N_{Cu,GB}$ and $N_{Ag,GB}$ are the numbers of Cu and Ag atoms in GBs, respectively, and $v_{Cu}$ and $v_{Ag}$ are the volumes of Cu and Ag atoms in respective single crystals. The GB atoms are identified using the common neighbor analysis algorithm. The excess free volume, defined by the numerator of the expression for $k$, represents the difference between the free volume of atoms in the GBs and the free volume of atoms in a perfect crystal. $k$ thus represents the normalized excess free volume per atom in GBs. Since there is typically an excess free volume in GBs (GBs are less dense than a perfect crystal), $k$ is equal or larger than zero. As shown in Fig.5b, the normalized excess free volume per atom $k$ in GBs decreases with increasing Ag concentration. The decrease in $k$ means that the GB is denser, which will makes GB atom jumps and free volume migrations more difficult. As a result, the decrease in $k$ by Ag dopants will drive up the resistance to GB sliding since less free volume is available to mediate GB sliding.

As shown above, Ag dopants decrease the free volume accessible to an individual GB atom, which process supports strengthening of the GBs. At the same time Ag doping increases the thickness of GBs (Fig.3b) and, since GBs are weaker than the crystalline grains, a larger volume fraction of GBs means that the alloys can become mechanically weaker. We propose that it is the balance of the two competing effects that is responsible for the presence of a maximum in the yield strength as a function of Ag concentration. In the small concentration regime, the GB sliding resistance increases and it is dominated by the effect of Ag on the free volume per atom in GBs. In the large concentration regime, GB sliding resistance decreases and it is dominated by the effects of Ag on the total volume fraction of GBs.

The above qualitative understanding can be quantified by fitting the yield stress (or equivalently the GB sliding resistance) vs. Ag concentration to the following relation $\sigma_{Yield} = \frac{A}{k^B \times N_{GB}{}^C}$. Here, $\sigma_{Yield}$ is the measured yield stress, $N_{GB}$ is the total number of GB atoms, and $A, B, C$ are the fitting parameters. The formula includes the effects of both, the free volume of GB atoms ($k$) and the number of GB atoms ($N_{GB}$), on the yield stress. The best numerical fits of the simulation data to this formula are shown in Fig.5a.

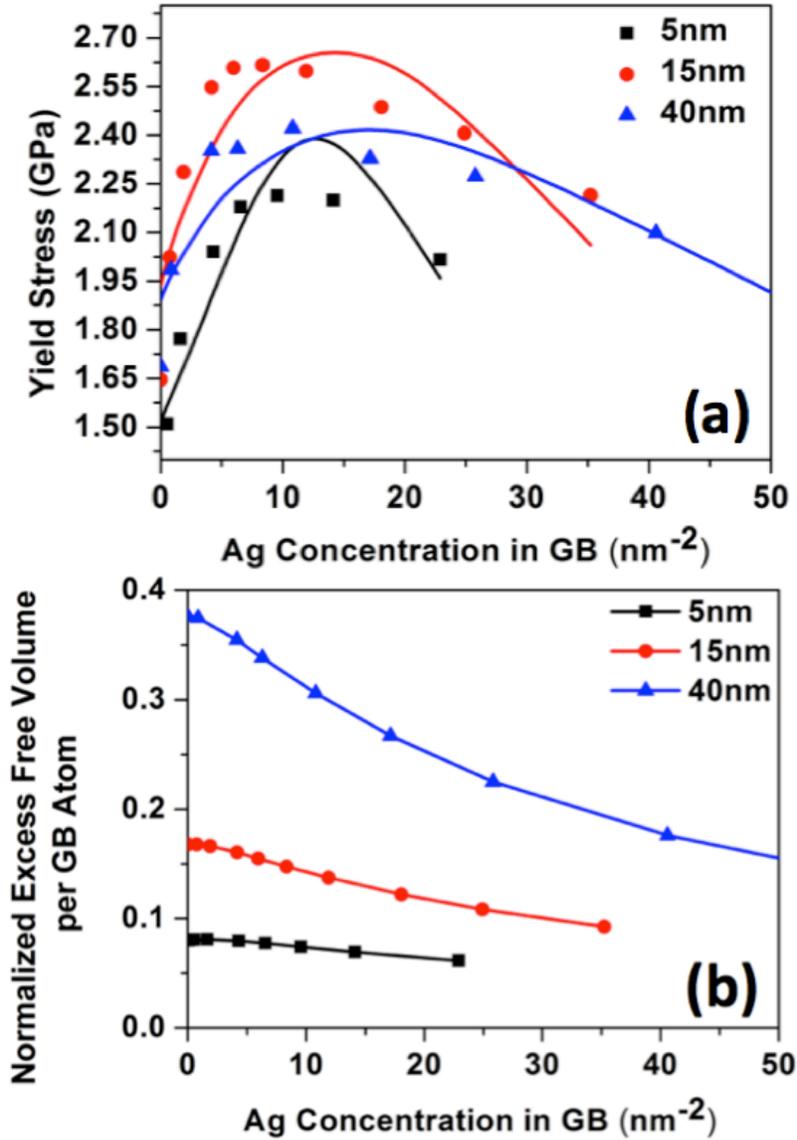

***Figure 5.*** *Dependence of the (a) yield stress (defined as 0.2%-offset) and (b) k, which is the normalized excess free volume per GB atom on Ag concentrations when the average grain diameters are 5, 15, and 40nm. The solid lines in (a) represent the best numerical fits of the simulation data to the formula* $\sigma_{Yield} = \frac{A}{k^B \times N_{GB}{}^C}$.

### 3.2.3 Flow Stress

In Fig.6a we show the dependence of the flow stress on the concentration of Ag for nc Cu/Ag alloys with the average grain diameters of 5nm, 15nm, and 40nm. Again, only those Ag concentrations are considered for which no wetting layers are formed. We find that the flow stress first increases and then decreases with the increase in Ag concentration. Moreover, for the 40nm grain diameter, the flow stress is less sensitive to Ag concentration than for the 5nm grain diameter. It is worth mentioning that for the case of zero Ag concentration, the flow stress increases when grain diameter increases from 5nm to 15nm, and then decreases when the grain diameter increases from 15nm to 40nm. This trend with increasing grain size is due to the transition from GB-plasticity to dislocation-mediated plasticity and this observation is consistent with published literature.[5,6,7]

As the flow stress is calculated in the strain regime where both GB sliding and intragranular dislocation plasticity are active, we analyze changes in these two mechanisms with Ag concentration. Resistance (or critical shear stress) to GB sliding is proportional to the yield stress (see Section 3.2.2) and the dependence of the yield stress on Ag concentration has been already plotted in Fig.5a. Two conclusions from the analysis of this plot that are worth reiterating here are: GB sliding resistance has a maximum for a finite Ag concentration and GB sliding resistance is less sensitive to Ag concentration for 40nm grain size than for 5nm grain size. Even though the yield stress (or GB sliding) calculations had been carried out before dislocations were nucleated, we assume that the same effect of Ag on GB sliding resistance is still present in the large strain regime, where dislocation plasticity is active.

To determine if alloying has any effect on the intragranular dislocation activity, we calculate dislocation density as a function of Ag concentration (see Fig.6b). We do not include GB dislocations in this analysis as our goal is to analyze mechanisms underlying the strength of crystalline grains. Also, the dominant GB deformation mechanism in the regime of Ag concentration that we study is migration of free volume and not glide of dislocations within GBs. Interestingly, we find that even though Ag is present only at GBs, Ag dopants significantly influence dislocations in the grain interiors. As shown in Fig.6b, dislocation density decreases with an increase in Ag concentration. Since in general the strength of a crystal is expected to increase with an increase in dislocation density due to dislocation interactions and entanglements, the results in Fig.6b imply that the strength of the grain interior (or resistance to intragranular deformation) decrease with increasing Ag concentration. There are multiple mechanisms by which Ag at GBs could influence dislocation activity inside the crystalline Cu grains and these mechanisms can be categorized as either direct or indirect. Direct effects can involve changes in dislocation-GB interactions, such as: dislocation nucleation at GBs, dislocation glide within confined grains (with the ends of dislocation lines remaining on GBs while the dislocation sweeps through a grain), and dislocations' transmission across GBs. All these phenomena could impact dislocation densities. The indirect effect of Ag dopants means that Ag dopants can influence GB sliding resistance and thus change the total contribution of dislocation-mediated plasticity. It is challenging to isolate the contributions of each of the above mechanisms based on simulations in a complex nc system. Further research is needed to investigate the effects of Ag dopants on each of these mechanisms using more simplified system geometries. Finally, it is also interesting that dislocation density becomes less sensitive to Ag concentration when grain diameter is as large as 40nm. The reason is that when the grain size increases, the volume fraction of the GB region decreases. In this case, the dislocation-GB interactions are less important to dislocation densities than the dislocation-dislocation interactions.

Given the above effects of Ag dopants on the GB sliding resistance and the dislocation slip resistance, we first discuss how the dependence of flow stress on Ag concentration reported in Fig.6a can be understood qualitatively. In the small concentration regime, the GB sliding resistance increases and the dislocation slip resistance decreases with the increase in Ag concentration. The increase in the flow stress thus indicates that the effect of Ag on GB sliding resistance plays a bigger role than Ag effect on dislocation slip resistance. In the large concentration regime, both the GB sliding resistance and the dislocation slip resistance decrease, leading to the decrease in the flow stress. In addition, the reason the flow stress in samples with grain sizes of 40nm or larger is less sensitive to Ag concentration than smaller-grain samples is that both the GB sliding resistance and the dislocation slip resistance are less influenced by Ag concentration in the large grain size regime.

In order to quantify the above analysis, we fit our simulation data in Fig.6a to the following formula for the flow stress: $\sigma_{Flow}(C_{Ag}) = k_1 \times \sigma_{Yield}(C_{Ag}) + [k_2 + k_3 \times \rho_{disl}(C_{Ag})]$. Here, $\sigma_{Yield}$ is the yield stress, $\rho_{disl}$ is the dislocation density, and $k_1$, $k_2$, and $k_3$ are fitting parameters. $\sigma_{Flow}$, $\sigma_{Yield}$, and $\rho_{disl}$ all depend on Ag concentration $C_{Ag}$. The first term, $k_1 \times \sigma_{Yield}(C_{Ag})$, represents the contribution from GB sliding resistance to the flow stress whereas the second term, $k_2 + k_3 \times \rho_{disl}(C_{Ag})$, represents the contribution from the dislocation slip resistance. The form of the dislocation-related term can be rationalized by considering the relation between dislocation-controlled strength and material's grain size. Specifically, when the grain size is large (larger than approximately 100nm), material's strength is related to the grain size $d$ via the Hall-Petch relation $\sigma_{dislocation} = \sigma_0 + c \times d^{-\frac{1}{2}}$,[32] where $c$ is an empirical parameter. For a smaller grain size, the power of $-\frac{1}{2}$ in the Hall-Petch relation gradually changes and material's strength is better described by the relation $\sigma_{dislocation} = \sigma_0 + c \times d$.[33,34] Moreover, for the same deformation strain, experimental data shows that dislocation density $\rho_{disl}$ is proportional to the inverse of the grain size $d \propto \frac{1}{\rho_{disl}}$.[35] This relation has also been found in our simulations (see Fig.6c). By plugging in this relation into the expression for strength, we get $\sigma_{dislocation} = k_2 + k_3 \times \rho_{disl}(C_{Ag})$, which is the same as the second term in our proposed expression for the flow stress $\sigma_{Flow}(C_{Ag})$. The fit of $\sigma_{Flow}(C_{Ag})$ to our simulation data is shown as solid lines in Fig.6a.

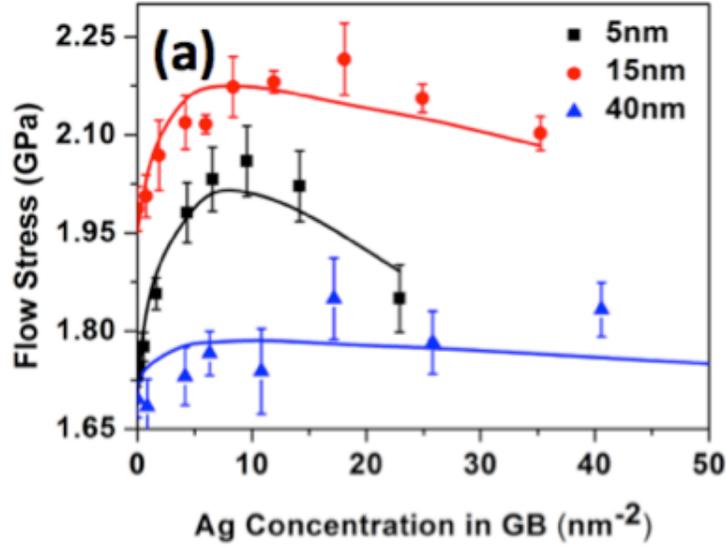

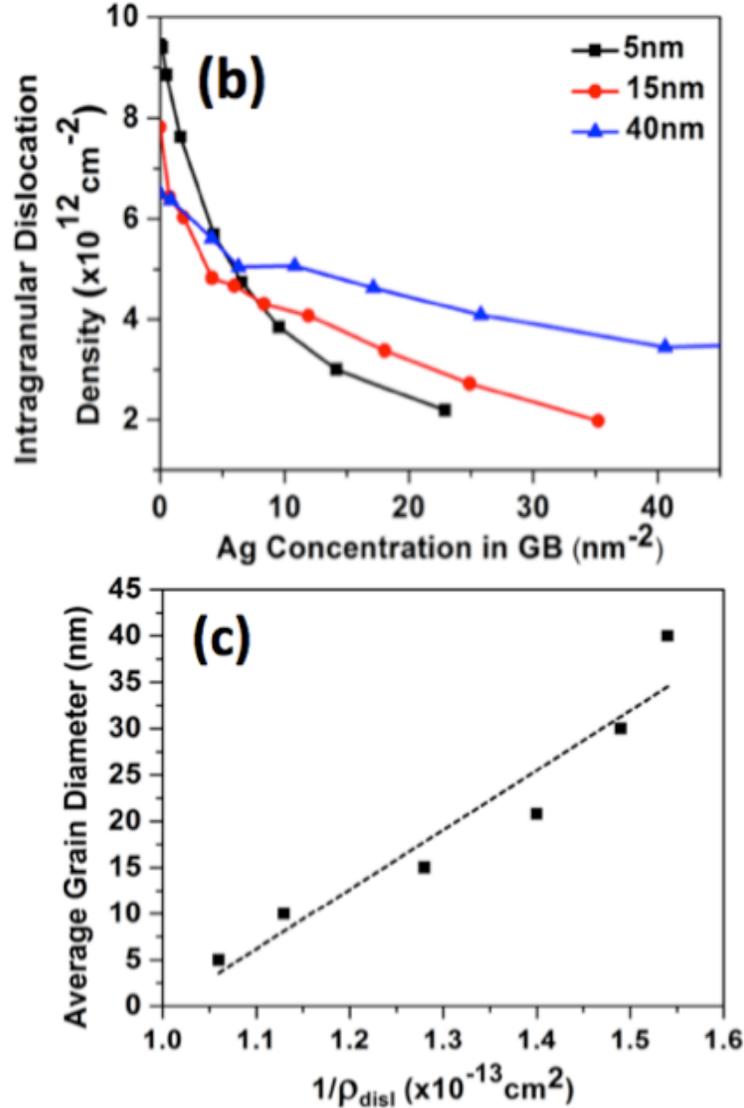

*Figure 6.* Dependence of (a) flow stress and (b) dislocation density on Ag concentration for samples with the average grain diameters of 5, 15, and 40nm. Solid lines in (a) are fits to $\sigma_{Flow}(C_{Ag}) = k_1 \times \sigma_{Yield}(C_{Ag}) + [k_2 + k_3 \times \rho(C_{Ag})]$. (c) Relation between the average grain diameter and the inverse of dislocation density. Solid line represents a linear fit. Data in (a), (b) and (c) corresponds to the engineering strain of 15%.

**4. Conclusions**

The microstructures and mechanical properties of Cu/Ag alloys have been discussed in this paper. When the concentration is below 50 Ag atoms/nm², Ag dopants at GBs of Cu/Ag alloys evolve from monolayer complexions to nanolayer complexions. Above the 50 Ag atoms/nm² threshold concentration, a crystalline Ag bulk phase is observed in a form of a wetting layer. Most of the interfaces between the Ag phase and the Cu grains are cube-on-cube and hetero-twin interfaces. The 0.2%-offset yield stress first increases and then decreases with an increase in Ag concentration, which is due to Ag dopants effects on GB free volume. The flow stress of the nc Cu/Ag alloys first increases and then decreases with an increase in Ag concentration, because Ag dopants are able to alter GB sliding resistance and dislocation slip resistance simultaneously. The above effects of Ag on the flow stress and deformation mechanisms are less significant when the average grain size is as large as 40nm.

**Acknowledgment**
The authors gratefully acknowledge financial support from Army Research Office Grant No. W911NF-12-1-0548.

**Conflict of Interest**
The authors declare that they have no conflict of interest.